# Trans-Coordinate States

By Richard A Mould

Department of Physics and Astronomy, State University of New York, Stony Brook, N.Y. 11794-3800.  http://ms.cc.sunysb.edu/~rmould

**Abstract**

*A meaning is given to the "state" of a system that is consistent with the trans-coordinate physics previously proposed by the author.  The collapse of such a state is described using the previously proposed q-rules.  A Hamiltonian is defined for these states and it is shown that their square modulus is conserved.  Keywords: invariance, state reduction, wave collapse.*

## Introduction

In previous papers it was shown that a particle's quantum mechanics wave function and its derivatives can be defined independent of space-time coordinate systems [1].  Differentials are obtained from a trans-coordinate limiting process that allows a dynamic principle to be applied *locally* everywhere within a particle's wave packet.  This provides for the evolution of the entire wave packet over the invariant metric manifold.  However, it is not possible in trans-coordinate physics to define the state of a system along a space-like plane (one for which *t* = constant) as is done in conventional coordinate-based physics.  A *non-planar definition of state* is proposed in this paper that discards the notion of coordinate-based simultaneity.

Although the dynamics is locally defined, the 'regional domain' of a wave function is altered *during a collapse*.  In order to preserve invariance, a wave collapse is transmitted through invariant metric space over the surface of the backward time cone of the initiating event (such as a measurement event) in the manner described by Hellwig and Kraus [2].  The original H-K collapse leads to troublesome causal loops [3], but this objection is not valid if a *modified* Hellwig-Kraus collapse is adopted that makes use of trans-coordinate states.  When that is done, the influence of a collapse is still transmitted over the surface of the backward time cone, but we make use of the q-rules developed in a previous paper to show that there are now *absolute causal priorities* that overcome the a-causal objections to a H-K collapse [4]. These causal priorities are shown to result in a unique Minkowski architecture among collapse events.

Also in this paper a Hamiltonian is defined for trans-coordinate states, and the rule is given for the conservation of expectation value of an operator *P*. A discontinuous interaction (like a quantum jump) is evaluated in the interaction picture and it is shown that the square modulus of a trans-coordinate state is conserved.

## Definition of state

In this physics, events are identified with non-systematic letters in the manner of Euclid. A wave equation of three particles is therefore of the form

$$\Psi(\mathbf{a}, \mathbf{b}, \mathbf{c}) = \psi_1(\mathbf{a})\psi_2(\mathbf{b})\psi_3(\mathbf{c})$$

where **a**, **b**, and **c** are three events that are confined to their respective wave functions and are required to have a space-like relationship to each other. There is no common time implicit in this equation because it is not defined over a single space-like surface to which a common time can be assigned. Each particle has its own space-time differentials defined by the trans-coordinate limiting process in Ref. 4, where integrated space-time variables are not introduced.

Consider a *coordinate-based* wave function with the spatial variables integrated out given by

$$\Psi(t_f \geq t \geq t_0) = a(t)b(t) + c(t)$$

This is like the case in which a free particle $a(t)$ interacts with a detector $b(t)$, where $c(t)$ is the same detector after capture. This way of representing the state assumes that there is a parameter *t* that "simultaneously" represents all three objects in the system. That is too much to assume in a trans-coordinate universe.

The same *trans-coordinate state* is given by

$$\Psi(\mathbf{a}, \mathbf{b}, \mathbf{c}) = a(\mathbf{a})b(\mathbf{b}) + \underline{c}(\mathbf{c}) \tag{1}$$

where **a** is any event inside the free particle wave packet, event **b** lies inside the pre-capture detector, and **c** is any event in the post-capture detector. All three events occur between the initially given state of the system and the final moment of capture, and all three are required to have a space-like relationship to each other. There are obviously an infinite number of ways to represent this state, corresponding to the infinite number of events **a**, **b**, and **c** that might be chosen for that purpose.

## Entangled particles

The zero spin state of two spin-entangled particles $p_1$ and $p_2$ at time $t$ is normally given by

$$\Phi(t) = \{p_1(t,\uparrow)p_2(t,\downarrow) - p_1(t,\downarrow)p_2(t,\uparrow)\}/\sqrt{2}$$

In the trans-coordinate case this is written

$$\Phi(\mathbf{a},\mathbf{b}) = \{p_1(\mathbf{a}\uparrow)p_2(\mathbf{b}\downarrow) - p_1(\mathbf{a}\downarrow)p_2(\mathbf{b}\uparrow)\}/\sqrt{2}$$

This particle state is taken together with two spin detectors, $d_1$ at event **m** and $d_2$ at event **n**, to make up the initial state of the system given by

$$\Psi = \Phi(\mathbf{a},\mathbf{b})d_1(\mathbf{m})d_2(\mathbf{n}) \qquad (2)$$

This equation is pictured in Fig. 1a. The world lines of the detectors are labeled $d_1$ and $d_2$, where **m** and **n** are space-like separated. The locations of **m** and **n** occur *prior to* the interaction – when a particle's world line overlaps its detector's world line). The world lines of the particles are the slanted shaded lines labeled $p_1$ and $p_1$

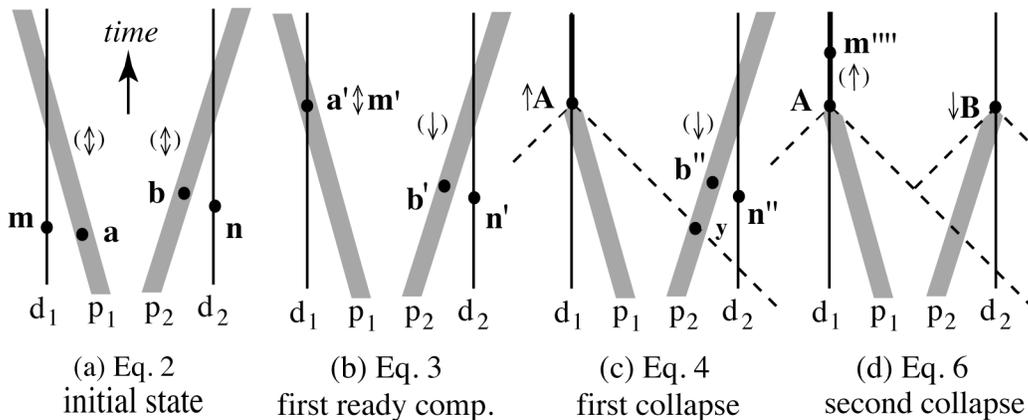

(a) Eq. 2          (b) Eq. 3              (c) Eq. 4            (d) Eq. 6
initial state    first ready comp.    first collapse     second collapse that are identified by events **a** and **b**. Their location along these world lines occur prior to their interaction with the detectors, and all four events are space-like separated. Again, there may be an infinite number of ways that this state can be represented, corresponding to the infinite number of ways that events **a**, **b**, **m**, and **n** might be chosen.

Figure 1: Corrolated particle measurement

The double vertical arrows in Fig. 1a indicate that each particle is in a superposition of spin-up and spin-down. The diagrams in Fig. 1 lack trans-coordinate significance because they are specific to the displayed Lorentz frame. However, Eq. 2 and all the other q-rule equations given below are completely trans-coordinate.

There are six possible outcomes of an interaction between one or both of these particles with their detectors. These appear in the form of six 'ready' components shown in the right-hand column of Eq. 3 that are generated by the dynamic principle acting on the initial state (i.e., the state in Eq. 2 that is the first component in Eq. 3).

$$\Psi + \underline{\Psi}' = \Phi(\mathbf{a}, \mathbf{b})d_1(\mathbf{m})d_2(\mathbf{n}) + \underline{d}_1(\mathbf{a'}\uparrow\mathbf{m'})p_2(\mathbf{b'}\downarrow)d_2(\mathbf{n'}) \qquad (3)$$
$$+ \underline{d}_1(\mathbf{a'}\downarrow\mathbf{m'})p_2(\mathbf{b'}\uparrow)d_2(\mathbf{n'})$$
$$+ \underline{d}_2(\mathbf{b'}\uparrow\mathbf{n'})p_1(\mathbf{a'}\downarrow)d_1(\mathbf{m'})$$
$$+ \underline{d}_2(\mathbf{b'}\downarrow\mathbf{n'})p_1(\mathbf{a'}\uparrow)d_1(\mathbf{m'})$$
$$+ \underline{d}_1(\mathbf{a'}\uparrow\mathbf{m'})d_2(\mathbf{b'}\downarrow\mathbf{n'})$$
$$+ \underline{d}_1(\mathbf{a'}\downarrow\mathbf{m'})\,d_2(\mathbf{b'}\uparrow\mathbf{n'})$$

The initial state is a *realized component*, which means that it is empirically real as explained in Ref. 4. The six outcome components on the right are called *ready components* that are identified by the underline of their first state. They are *not* empirically real as is also explained in Ref. 4. During the interaction the probability current flows from the initial component in Eq. 3 into some combination of the six ready components, establishing the probability that one of them will be stochastically chosen during a time interval *dt*. When that happens the chosen component will become realized and all the others will collapse to zero.

The state $\underline{d}_1(\mathbf{a'}\uparrow\mathbf{m'})$ in the first ready component of Eq. 3 represents the interaction of the 1st particle with its detector with its spin up; whereas $p_2(\mathbf{b'}\downarrow)$ in that component represents the non-interacting 2$^{nd}$ particle with its spin down. The second ready component begins with the 1$^{st}$ particle interacting with its spin down $\underline{d}_1(\mathbf{a'}\downarrow\mathbf{m'})$ and the non-interacting 2$^{nd}$ particle with its spin up $p_2(\mathbf{b'}\uparrow)$. The third and fourth ready components describe an interaction of the 2$^{nd}$ particle with similar spin down and spin up possibilities. The fifth and sixth ready components in Eq. 3 represent a dual interaction that leads to the possibility that both particles will be stochastically chosen "together". This is highly improbable.

There are three possible scenarios. The first is that probability current from the realized component will flow into both the first and second ready components, but not into the other four. This will happen when the 1$^{st}$ particle interacts with its detector but the 2$^{nd}$ particle does not. The second scenario is that probability current will flow from the realized component into the third and fourth ready components, but not into the other four. This will happen when the 2$^{nd}$ particle interacts with its detector but the 1$^{st}$ particle does not. The third scenario is that probability current flows 'at once' into all six ready components. This happens when the 1$^{st}$ and 2$^{nd}$ particles interact 'together' with their separate detectors.

Figure 1b is a diagram of the first scenario consisting of the first and second ready components in Eq. 3. We cannot include all of Eq. 3 on a single diagram, so we represent just one possible scenario. It shows the 1$^{st}$ particle interacting with the first detector with spin still undetermined. The 2$^{nd}$ particle is still not interacting with the second detector in that diagram. Figure 1b is represented by the equation

$$\Psi + \underline{\Psi}' = \Phi(\mathbf{a}, \mathbf{b})d_1(\mathbf{m})d_2(\mathbf{n}) + \underline{d}_1(\mathbf{a'}\uparrow\mathbf{m'})p_2(\mathbf{b'}\downarrow)d_2(\mathbf{n'})$$
$$+ \underline{d}_1(\mathbf{a'}\downarrow\mathbf{m'})p_2(\mathbf{b'}\uparrow)d_2(\mathbf{n'})$$

in which the position of event **a** (the particle) is understood to coincide with that of the event **m** (the detector). Probability current flows equally into both of the ready components in this equation; so if the first of these components is statistically chosen it will become a realized component, and the other two components will go to zero. The measured state of the system is then given by

$$\Psi'' = d_1(\uparrow\mathbf{A})p_2(\mathbf{b}''\downarrow)d_2(\mathbf{n}'') \tag{4}$$

where $d_1(\uparrow\mathbf{A})$ is event **A** in Fig. 1c.  This is the vertex of a modified Hellwig Kraus collapse.  Above the backward time cone of event **A** the state is identical with that shown in Fig. 1b.  The important thing about event **b''** in Fig. 1c is that it occurs causally 'after' event **y** (on the surface of the backward cone) and causally 'before' interacting with its detector.  The darkened world line above event **A** refers to the first detector that now includes an additional particle.

We assume that the 2$^{nd}$ particle will subsequently interact with its detector, so Eq. 4 generates the ready state $\underline{\Psi}'''$ in Eq. 5 that is *not* diagramed in Fig. 1

$$\Psi''' + \underline{\Psi}''' = d_1(\uparrow\mathbf{A})\,p_2(\mathbf{b}''\downarrow)d_2(\mathbf{n}'') + \underline{d}_1(\uparrow\mathbf{m}''')d_2(\mathbf{b}'''\downarrow\mathbf{n}''') \tag{5}$$

When the ready component in this equation is stochastically chosen it collapses to

$$\Psi'''' = d_1(\uparrow\mathbf{m}'''')d_2(\downarrow\mathbf{B}) \tag{6}$$

defining event **B** in Fig. 1d.  And finally

$$\Psi''''' = d_1(\uparrow\mathbf{m}''''')d_2(\downarrow\mathbf{n}''''') \tag{7}$$

that is not diagramed in Fig. 1.

The collapse in Eq. 4 or 6 is assumed to be instantaneous.  It follows from the rules of collapse given in Ref. 4.  The transition time $\Delta T$ is ignored in all of these cases.

**The causal order**

Event **A** occurs 'absolutely' before event **B** because the first ready eigenstate is chosen in Eq. 3 – excluding other choices.  Once that is done, there is no going back to the other choices.  Event **B** cannot influence event **A** because **A** has already happened in Eq. 4, which is a realized state *before* the 2$^{nd}$ particle interacts with its detector.  This is shown in Fig. 1d where the backward light path emanating to the left of event **B** does not penetrate the backward time cone of event **A**.  On the other hand, event **A** can influence event **B** as is apparent in Eqs. 4-6.  This asymmetry is implicit in the stochastic choice of one of the six possibilities in Eq. 3, and has nothing to do with the Lorentz frame that is used to picture the events.  The temporal order pictured in Fig. 1d imagines that **A** and **B** occur at the *same time*, but that is not the causal order determined by Eq. 4. The relativistic ordering of events as seen by different Lorentz observers has nothing to do with the absolute causal ordering that puts event **A** before event **B**. The relativity of simultaneity is a coordinate idea that is based on the way things appear in different Lorentz frames.

Even 'relative' simultaneity is a meaningless idea. There are only two ways that an event can be temporally related to another space-like separated event. (i) Both are inside of particles that are classically correlated or quantum mechanically entangled with one another through interactions that occurred in their past; and (ii) they both share the symmetry properties of a mutual metric manifold. For instance, if two particles are spatially separated in a non-relativistic inertial system we would say that their internal clocks "tick" at the same rate because of item (ii). We might even "set them" to read the same time using a synchronization procedure based on item (i). However, they would not really give us the *same time* in any fundamental sense – because there is no such thing. If you and I are spatially separated and in relative motion it is not valid for you to speak of my future as though it already exists, or to say that my present doesn't yet exist. My present exists when I say it does, and my future exists right after that. You and your coordinates have no say in the matter.

The influence of a Hellwig-Kraus collapse may seem to extend infinitely far into space and time but that is not true. The limited influence of event **B** in Fig. 1d is characteristic of what generally happens. Every state reduction follows many other state reductions that will limit its influence. This is shown in Fig. 2 where several state reductions are arranged according to their causal priority. They appear as a mountainous landscape where the mountain peak in the foreground is causally prior to a peak in the background. This means that the ones in the back are limited in their influence by the ones in the front. In a 2 + 1 space, the landscape will appear as a two dimensional superposition of mountaintops on a landscape of prior mountaintops. In the improbable case that both events **A** and **B** are chosen in a dual interaction, neither one would influence the interior of the backward time cone of the other, so we would have a double-peaked mountain.

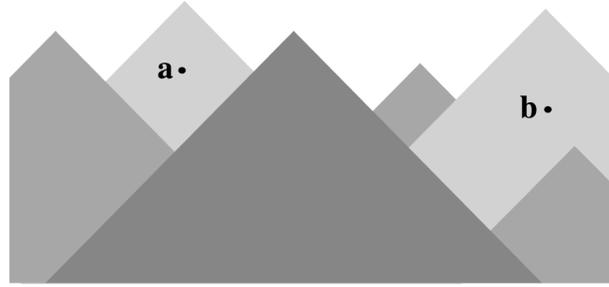

Figure 2: Mountainous architecture of Minkowski space

Every event in the universe is located in one or another of these mountainous peaks – like event **a** in the background peak on the left in Fig. 2, or event **b** in the middle-right peak. A modified Hellwig-Kraus collapse not only provides a non-local causal influence, it also organizes events into an absolute causal framework without regard to relativistic simultaneity. Figure 2 is the view from one Lorentz frame in which the foreground mountain peak is higher than the background peak immediately to its right. In another Lorentz frame the background frame might be higher than the foreground peak. However, the causal order will be the same in both. Every mountain peak has a space-like relationship to every other mountain peak.

**The Hamiltonian**

Let the wave function of a system of two *non-interacting* objects be given by

$$\Psi(\mathbf{a}, \mathbf{b}) = a(\mathbf{a})b(\mathbf{b})$$

Define the operator $d/dt_{12}$ through the trans-coordinate limiting process (Eq. 2 of Ref. 1) acting on $\Phi(\mathbf{a}, \mathbf{b})$ to be

$$\frac{d}{dt_{12}}\Psi(\mathbf{a}, \mathbf{b}) = \frac{d}{dt_1}\Psi(\mathbf{a}, \mathbf{b}) + \frac{d}{dt_2}\Psi(\mathbf{a}, \mathbf{b}) = \frac{da(\mathbf{a})}{dt_1}b(\mathbf{b}) + a(\mathbf{a})\frac{db(\mathbf{b})}{dt_2}$$

where $t_1$ is the time at event **a** and $t_2$ is the time at event **b**. Dropping references to **a** and **b**, the dynamic principle for these independent particles is therefore

$$i\frac{d}{dt_{12}}\Psi = H\Psi \quad \text{where } H = H_1 + H_2$$

$$i\frac{d}{dt_{12}}\Psi = i\frac{d}{dt_1}\Psi + i\frac{d}{dt_2}\Psi = H_1\Psi + H_2\Psi$$

giving
$$i\frac{d}{dt_1}\Psi = H_1\Psi \qquad i\frac{d}{dt_2}\Psi = H_2\Psi$$

or
$$i\frac{d}{dt_1}a = H_1 a \qquad i\frac{d}{dt_2}b = H_2 b \tag{8}$$

The expectation value of an operator $P(\mathbf{a})$ at an event $\mathbf{a}$ in the object $a(\mathbf{a})$ evolves in the usual way, except we leave it as per unit volume.

$$\frac{d}{dt} a^*Pa = (\frac{d}{dt}a^*)Pa + a^*P(\frac{d}{dt}a) + a^*\frac{\partial P}{\partial t}a$$

$$\frac{d}{dt} a^*Pa = i a^*[HP - PH]a + a^*\frac{\partial P}{\partial t}a$$

When $P$ commutes with $H$ and is explicitly time independent, its expectation values will be unchanged in time. Because the direction of the differential variable $dt$ through event $\mathbf{a}$ follows the square modular flow, the same may be said of the operator's expectation value per unit volume at event $\mathbf{a}$.

## A discontinuous interaction

Some interactions, like some scattering interactions, give rise to a continuous change in the wave function that occurs entirely within a single component of the wave function. Other interactions produce a discontinuous change that requires the creation of a new component that cannot arise continuously from the given component. If for instance there is a discontinuous interaction between two objects $a(\mathbf{a})$ and $b(\mathbf{b})$ giving rise to a new *ready* component $\underline{c}(\mathbf{c})$, then the state function $\Psi(\mathbf{a}, \mathbf{b}, \mathbf{c}) = a(\mathbf{a})b(\mathbf{b}) + \underline{c}(\mathbf{c})$ will be dependent on three differential times $dt_1$, $dt_2$, and $dt_3$ such that

$$\frac{d}{dt_{123}}\Psi(\mathbf{a}, \mathbf{b}, \mathbf{c}) = \frac{d}{dt_1}\Psi(\mathbf{a}, \mathbf{b}, \mathbf{c}) + \frac{d}{dt_2}\Psi(\mathbf{a}, \mathbf{b}, \mathbf{c}) + \frac{d}{dt_3}\Psi(\mathbf{a}, \mathbf{b}, \mathbf{c})$$

$$= \frac{da(\mathbf{a})}{dt_1}b(\mathbf{b}) + a(\mathbf{a})\frac{db(\mathbf{b})}{dt_2} + \frac{d\underline{c}(\mathbf{c})}{dt_3}$$

Again dropping references to **a** and **b**, the dynamic principle is now given by

$$i\frac{d}{dt_{123}}\Psi = H\Psi \quad \text{where: } H = H_1 + H_2 + \mathcal{H} \tag{9}$$

where $\mathcal{H}$ is the interaction Hamiltonian. In this case the best way to proceed is in the interaction picture.

## Interaction picture

The interaction picture introduces a new state variable

$$\Phi = \exp i(H_1 t_1 + H_2 t_2)\Psi$$

This function contains the integrated variables $t_1$ and $t_2$. One can take the view that these are not coordinates because they represent an integration of $dt_1$ and $dt_2$ along the separate world lines of objects 1 and 2. However these variables do imply a choice of origins for $t_1$ and $t_2$, and that makes them essentially coordinates. We therefore say that nature only goes so far as Eq. 9. That is entirely sufficient. That equation contains everything that is important. The function $\Phi$ is introduced as an analytic tool only. The entire interaction picture is merely analytic because it uses time coordinates.

The time derivative of $\Phi$ is given by

$$i\frac{d}{dt_{123}}\Phi = -H_1\Phi - H_2\Phi + i\exp i(H_1 t_1 + H_2 t_2)\frac{d}{dt_{123}}\Psi$$

$$= -H_1\Phi - H_2\Phi + \exp i(H_1 t_1 + H_2 t_2)(H_1 + H_2 + \mathcal{H})\Psi \quad \text{from Eq. 9}$$

$$= -H_1\Phi - H_2\Phi + (H_1 + H_2)\Phi + \exp i(H_1 t_1 + H_2 t_2)\mathcal{H}\Psi$$

giving $$i\frac{d}{dt_{123}}\Phi = \mathcal{H}_{\text{int}}\Phi \tag{10}$$

where $\mathcal{H}_{\text{int}} = \exp i(H_1 t_1 + H_2 t_2)\mathcal{H}\exp -i(H_1 t_1 + H_2 t_2)$

The new solution $\Phi$ is therefore propagated through time by the transformed interaction Hamiltonian in this picture.

**Square modular conservation**

From the fact that $\Phi = \exp i(H_1 t_1 + H_2 t_2)\Psi$ in the interaction picture, we have

$$\Phi^*\Phi = \Psi^*\Psi \tag{11}$$

Take

$$\frac{d}{dt_{123}} \Phi^*\Phi = \left(\frac{d}{dt_{123}} \Phi^*\right)\Phi + \Phi^*\left(\frac{d}{dt_{123}} \Phi\right)$$

and from Eq. 10

$$\frac{d}{dt_{123}} \Phi^*\Phi = i(\mathcal{H}_{int}\Phi)^*\Phi - i\Phi^*\mathcal{H}_{int}\Phi = 0 \tag{12}$$

inasmuch as $\mathcal{H}_{int}$ is Hermitean.

The square modulus of a q-rule component like $u(\mathbf{u}, t)$ at event $\mathbf{u}$ at a time $t$ is given by $u(\mathbf{u}, t)^*u(\mathbf{u}, t)$ according to the 'local' definition of square modulus in Ref. 1; so a component involving products like $u(\mathbf{u}, t_1)v(\mathbf{v}, t_2)$ has a square modulus equal to $u(\mathbf{u}, t_1)^*v(\mathbf{v}, t_2)^*u(\mathbf{u}, t_1)v(\mathbf{v}, t_2)$.

A q-rule equation like $u(\mathbf{u}..., t_1...) + v(\mathbf{v}..., t_2...)$ is defined in Ref. 2 in such a way that there must be a different composition of particles in one component than in the other; so one component has more particles, or less particles, or just different particles than the other component. Otherwise there would be only one component containing a complete description of the particle or particles involved. We will say that this difference in composition amounts to a trans-coordinate requirement of *orthogonality* between them. Therefore

$$[u(\mathbf{u}..., t_1...) + v(\mathbf{v}..., t_2...)]^*[u(\mathbf{u}..., t_1...) + v(\mathbf{v}..., t_2...)]$$
$$= u(\mathbf{u}..., t_1...)^*u(\mathbf{u}..., t_1...) + v(\mathbf{v}..., t_2...)^*v(\mathbf{v}..., t_2...)$$

If $u$ is a realized component and $v$ is a ready component, we may further generalize by saying that there can be no cross term between an empirically real component and an unreal one.

Combining Eqs. 11 and 12 with the function Ψ(**a**, **b**, **c**) = $a$(**a**)$b$(**b**) + $\underline{c}$(**c**) gives

$$\frac{d}{dt_{123}} \Psi^*\Psi = \frac{d}{dt_{12}} (ab)^*(ab) + \frac{d}{dt_3} \underline{c}^*c = 0 \qquad (13)$$

This tells us that the square modulus of Ψ is conserved in time when summing over each of the specified events **a**, **b**, and **c**. We arrive at this conclusion by using the *improper* function Φ in an intermediate analytic step, but that does not invalidate the conclusion. The three times $t_{123}$ are kept separate at every step of the way. Equation 13 is entirely trans-coordinate.